\documentclass[10pt,twoside,preprint]{elsarticle}

\usepackage{amssymb}
\usepackage{amsmath}

\usepackage{subfigure}
\usepackage{graphicx}
\usepackage{caption2}

\usepackage{algorithm} 
\usepackage{algorithmic} 
\usepackage{multirow} 
\usepackage{amsmath}
\usepackage{xcolor}
\usepackage{float}

\usepackage{amssymb}
\usepackage{amsthm}
\newtheorem{theorem}{Theorem}

\theoremstyle{definition}
\newtheorem{definition}[theorem]{Definition}
\theoremstyle{remark}

\theoremstyle{problem}
\newtheorem{problem}{Problem}
\theoremstyle{property}
\newtheorem{property}{Property}

\usepackage{geometry}   
\begin{document}
\begin{frontmatter}

\title{Full-scale Cascade Dynamics Prediction with a Local-First Approach}


\author[rvt,focal,oth]{Tao Wu\corref{cor1}}
\ead{wutaoadeny@gmail.com}
\author[rvt,focal,oth]{Leiting Chen}
\author[els]{Xingping Xian }
\author[rvt]{Yuxiao Guo}
\cortext[cor1]{Corresponding author}


\address[rvt]{Department of Computer Science and Engineering, University of Electronic Science and Technology of China}
\address[focal]{Institute of Electronic and Information Engineering in Dongguan, University of Electronic Science and Technology of China}
\address[oth]{Digital Media Technology Key Laboratory of Sichuan Province}
\address[els]{Department of Computer Science and Technology, Chengdu Neusoft University}

\begin{abstract}
Information cascades are ubiquitous in various social networking web
sites. What mechanisms drive information diffuse in the networks?
How does the structure and size of the cascades evolve in time? When
and which users will adopt a certain message? Approaching these
questions can considerably deepen our understanding about
information cascades and facilitate various vital applications,
including viral marketing, rumor prevention and even link
prediction. Most previous works focus only on the final cascade size
prediction. Meanwhile, they are always cascade graph dependent
methods, which make them towards large cascades prediction and lead
to the criticism that cascades may only be predictable after they
have already grown large. In this paper, we study a fundamental
problem: full-scale cascade dynamics prediction. That is, how to
predict when and which users are activated at any time point of a
cascading process. Here we propose a unified framework, FScaleCP, to
solve the problem. Given history cascades, we first model the local
spreading behaviors as a classification problem. Through data-driven
learning, we recognize the common patterns by measuring the driving
mechanisms of cascade dynamics. After that we present an intuitive
asynchronous propagation method for full-scale cascade dynamics
prediction by effectively aggregating the local spreading behaviors.
Extensive experiments on social network data set suggest that the
proposed method performs noticeably better than other
state-of-the-art baselines.
\end{abstract}


\end{frontmatter}

\section{Introduction}\indent

Online social networks provide platforms for people in where they
can easily share and discuss ideas and innovations. In this setting,
people react to information on the basis of their neighbors'
behavior, and the people in contact with them act in a same way.
Thus information cascades naturally form and become common in online
social networks. In consideration of the impact of information
cascades on online social networks, uncovering how information
cascades in the networks can considerably deepen our understanding
about information cascades and facilitate various vital
applications, including viral marketing, spreading suppression and
even link prediction. A growing body of researches has focused on
the statistical properties of information cascades or the common
patterns in temporal dynamics [1-5]. As a nontrivial line of work,
information cascades prediction has aroused considerable research
interests recently. Traditional models (independent cascade or
threshold model) are usually designed for all diffusion processes,
regardless of the nature of the diffusion objects. However, recent
studies [1, 2] show that the assumption that most of the models
followed becomes unreasonable for information cascades in online
social networks, and the information cascades, as the fundamental
collective dynamics of social networks, have strong relevance to
complex contagion mechanisms.

Availability of large scale data about information cascades has
facilitated the study of predictive models, and many approaches have
been developed recently, including model-centric methods [6-8] and
empirical methods [9-14]. Most of the previous methods try to
distinguish between messages with different popularity and assume
that the cascade graph (i.e. the path and user information of
information cascades.) in early stage is available. They are biased
towards studying extremely large but also extremely rare cascades,
by passing the whole issue about the general predictability of
cascades. However, because very limited cascade information can be
obtained for newly created cascades and storing complete cascade
graph may not be feasible as the size of cascades grows, the cascade
graph based methods may be impracticable in real-life applications.
Moreover, the real-life applications always care about not solely
the final cascade size, but also the whole issue of information
cascades. These challenges reinforce the fact that finding an
effective way to predict when and which users will be activated at
any time point of the entire lifetime of cascades regardless of the
cascade graph size remains a big task to date.

The proposed full-scale cascade dynamics prediction is different
from the traditional cascades prediction problems. Except for the
difference between the core aims of the proposed problem and the
traditional cascades prediction, as discussed above, the overriding
concern of the proposed problem is to evaluate all possible factors
and measure principal driving mechanisms by capturing the effective
information and the intrinsic relations between them, while the
focus of the classical problem is the design of discriminative
features. Moreover, the proposed problem calls for a common
computational framework that is independent of data features, which
is much more general than the traditional problems designed for
certain types of information in various real-life networks.

The full-scale cascade dynamics prediction presents several
challenges. First, hybridity, information cascades are driven by the
social interactions between users with various confounding factors,
which makes it difficult to be modeled comprehensively and
effectively. That is, various factors are considered in previous
methods [9, 10, 11, 14], and how to comprehensively study and
identify the effective factors is a big challenge. Second,
incomplete, in this paper, we argue that getting global scenes of
information flows is usually unfeasible, which makes the problem
nontrivial. Therefore, how to effectively capture the whole issue of
information cascades with little global cascade information becomes
a significant challenge. Third, generality, it is important to
develop a model that can adapt well to different types of social
media, including blogging, e-mail, social sites, etc.


In light of these differences and challenges, we define ``spreading
behaviors" to represent the personal interactions and behaviors at
microscopic level and assume that the macroscopic information
cascades are generated from the spreading behaviors at microscopic
level. Thus, the task of full-scale cascade dynamics prediction can
be decomposed into a set of spreading behavior estimation tasks with
network topology, and the availability of cascade graphs in early
stage is no longer the prerequisite for information cascades
prediction. Based on the above analysis, this paper proposes a
unified two-phase framework FScaleCP to predict full-scale cascade
dynamics. At the first phase, it leverages supervised learning to
model local spreading behaviors by capturing the joint action of
factors from multiple dimensions and quantitatively selects the most
effective feature space and spreading behavior estimation model. At
the second phase, Inspired by the work on label propagation
algorithm [15], we incorporate the spreading behavior estimation
model into local network topology and propose a flexible and robust
supervised asynchronous propagation method to aggregate the local
spreading behaviors for full-scale cascade dynamics prediction.
Figure 1 gives an illustration of information cascade dynamics
prediction. The main contributions of this work include: (1) we
formally define a novel problem of full-scale cascade dynamics
prediction in networks and propose a unified FScaleCP framework to
solve it; (2) we propose a supervised asynchronous propagation
method to incorporate the local spreading behaviors for predicting
cascade dynamics in social network. To estimate the local spreading
behaviors, we adopt the supervised learning model and feature
selection method to utilize implicit and confounding factors; (3)
experimental results on social network data set demonstrate that the
proposed FScaleCP significantly outperforms several state-of-the-art
information cascades prediction algorithms. Together, these
contributions would promote the understanding of how information
cascades across network and suggest new directions for modeling and
optimization of information cascades.

The remainder of this paper is organized as follows. Section 2
presents related works about information cascades. Section 3
formalizes the full-scale cascade dynamics prediction problem in
social networks. Section 4 details the proposed framework. Section 5
explains the experimental results. Section 6 concludes the work.

\begin{figure}[!h]
\centering \subfigure[]{ \label{fig:side:a}
\includegraphics[width=2.5in]{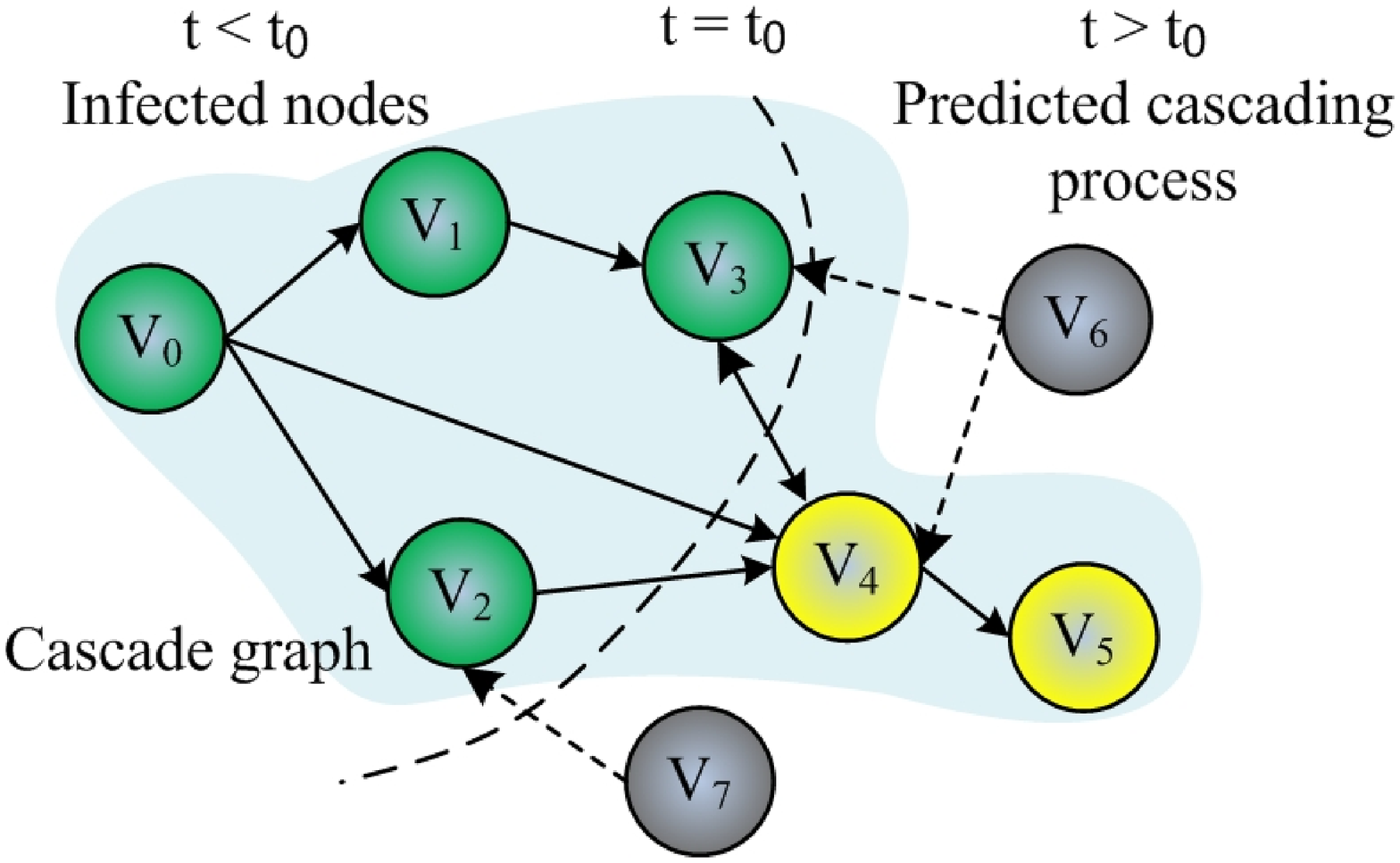}}
\hspace{8ex} \subfigure[]{ \label{fig:side:b}
\includegraphics[width=3.3in]{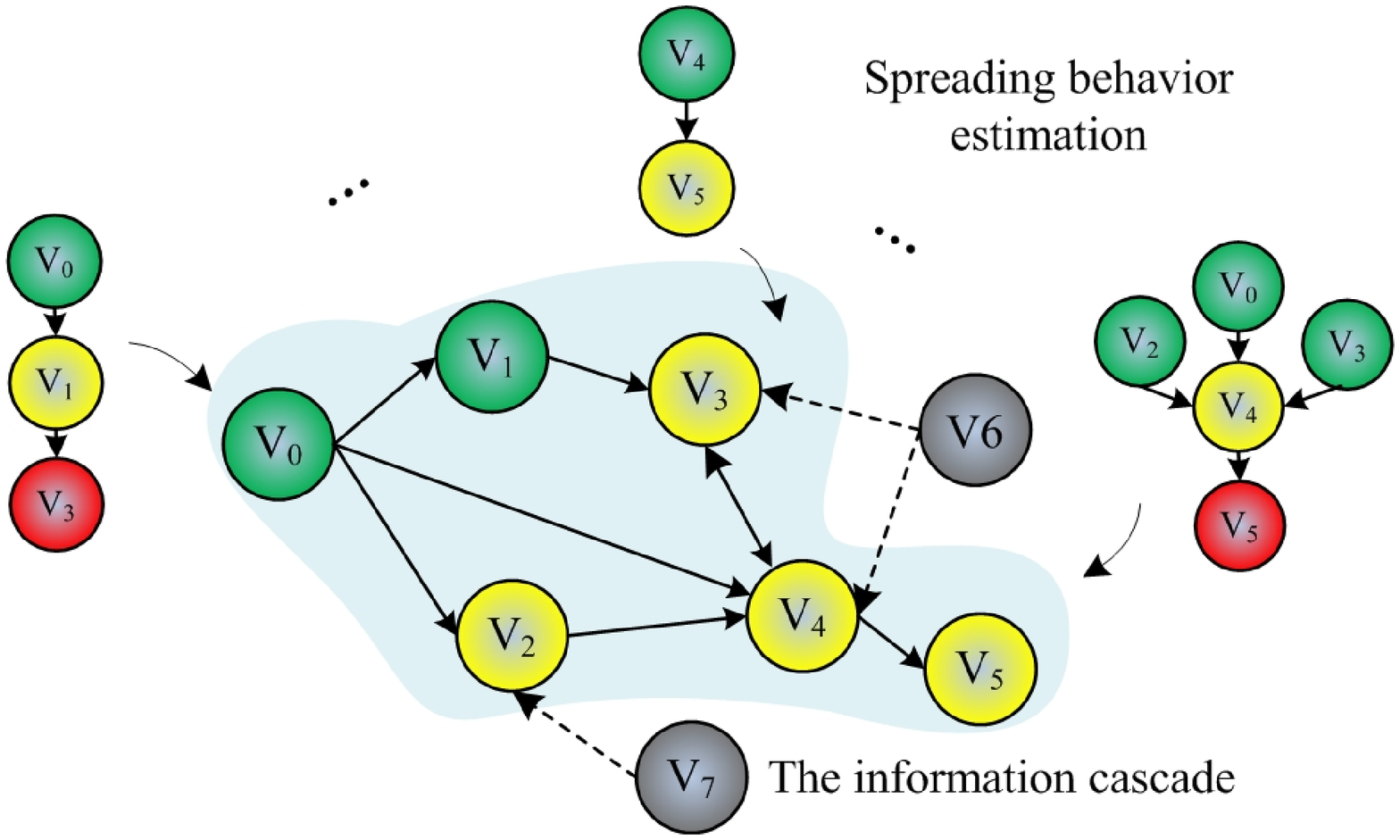}}
\caption{Illustrate of information cascade dynamics prediction.
 (a) Cascading Process; (b) Behavioral dynamics estimation based
cascade dynamics prediction, in which nodes denote network users and
dashed lines represent inter-relationships. In the network, the
evolution of information flows (solid lines) is predicted based on
atomic spreading behavior estimation tasks. Meanwhile, all nodes
update their network behaviors asynchronously on the basis of the
actions of their surroundings. }\label{fig:side}
\vspace{\baselineskip}
\end{figure}

\section{Related work}\indent
In this section we review some important researches that are close
related to our work, regarding social influence computation,
individual spreading behavior estimation, as well as information
cascades prediction problem.

\textbf{Social influence.} Social influence accompanies transferring
information from one user to the other. It is a key to explain
information cascades in social networks. A crucial task in the
analysis of social influence is to find evidence of influence and
distinguish influence from homophily or unobserved confounding
variables. Shuffle test [16] is proposed to detect a signal of
influence based on the intuition that if influence is not a likely
source of correlation in a system, timing of actions should not
matter, and therefore reshuffling the time stamps of the actions
should not significantly change the amount of correlation.
Investigations on the interplay between social influence and
homophily using data from Wikipedia have been made by Crandallet et
al. [17]. S. Aral et al. [18] develop a dynamic matched sample
estimation framework to distinguish influence and homophily effects
in dynamic networks. Moreover, all viral marketing papers assume
they have a social graph with edges labeled with probabilities of
influence between users as input. To our knowledge, the question of
how or from where one can compute the influence probabilities has
been largely left open. Amit Goyal et al. [19] study how to learn
cascade probabilities from a log of past propagations, and they
propose both static and time-dependent models for capturing
influence. Tang et al. [20] propose a topic factor graph model to
measure the strength of topic-level social influence quantitatively.

\textbf{Individual spreading estimation.} A bulk of studies attempt
to understand individual spreading behaviors in terms of decay
effects [21], memory effect [22], social reinforcement [22, 23],
etc. Moreover, some other works show that limited attention [24, 25]
and semantic similarity [24] are also important indicators. With
limited attention, information is less likely to be noticed by users
as the volume of information scales with the number of user's
parents. Meanwhile, users tend to spread the information with
similar semantic meaning to their history information. Besides, Tang
et al. [26] study the effects of pairwise influence and structure
influence on user's spreading behaviors. As the focus of the
existing works is to explore whether a factor is effective in
spreading behavior estimation, rather than modeling spreading
behavior as accurate as possible, the prediction accuracy they
produce is less reliable and there is no knowledge about the
importance of every factor.

\textbf{Information cascades prediction.} Information cascades
prediction has broad application range, including item
recommendation, viral marketing, spreading suppression, etc. It is
the research area that is most relevant to our work. Many
theoretical model based methods are developed to capture the
cascading process [6-8]. As the complexity limitation and many
assumptions being included, the methods may hard to consider the
complex contagion mechanisms comprehensively, and their solutions
are generally not applicable in real-life situations. On the other
hand, to predict the future popularity of a meme, Weng et al. [9]
develop a model considering early cascade patterns based community
concentration, influence of early adopters and time series
characteristics. Ma et al. [10] propose a supervised learning based
method to predict the range of the popularity of new hashtags in
Twitter. It extracts content features from hashtag string and the
collection of tweets and contextual features from cascade graph.
Hong et al. [11] formulate the task of predicting the popularity of
messages into a multi-class classification problem by investigating
content features, temporal features, as well as structural
properties of cascade graph. Kupavskii et al. [12] try to forecast
how many retweets a given tweet will gain during a fixed time period
since the initial moment. Tsur et al. [13] also study the diffusion
of information in Twitter and predict the hashtag popularity by
combining content features with temporal and topological features.
Cheng et al. [14] focus on a some different problem. Rather than to
predict the final cascade size, they instead to predict whether the
cascade size of the next stage will bigger than the median site and
seek to understand how predictability varies along the entire
lifetime of cascade. They are aware of the dependency of early
cascade graphs and the bias of extremely large cascades, but they
don't solve the problem thoroughly.

\textbf{Summary.} To the best of our knowledge, our work is the
front-runner to propose that modeling full-scale cascade dynamics
indirectly using local spreading behavior estimation, which optimize
the generality of cascade prediction method and capture the whole
issue of information cascades. This distinguishes our work from the
existing studies that merely consider the final size prediction of
extremely large but also extremely rare cascades.

\section{Problem formulation}\indent

Generally, we use $ G = (V; E) $ to denote a social network, where $
V = \{v_i\} $ is the set of nodes which represent social users, and
$ E \subseteq V \times V $ is the set of directed edges which are
mapped to links between social users. $e_{ij} = (v_i, v_j ) \in E$
means there is a direct link from node $v_i$ to $v_j$ in $ G $ and
$e_{ij} \notin E$ represents that there is no such a direct link. To
a directed edge $e_{ij}$, we define node $v_j$ is the parent of node
$v_i$ and node $v_i$ is the child of node $v_j$, and $v_i$ is
exposed to the messages published by $v_j$. Then we define $P_i$ to
represent the parent node set of node $v_i$ and $C_i$ to represent
the child node set of node $v_i$. Moreover, if node $v_j$ is a
parent of node $v_i$ and node $v_i$ is also a parent of node $v_j$,
we say they are friends.

\begin{definition} [Activated nodes]
Given node $v_j$ and its child node $v_i$, if node $v_j$'s action
associated to a message induce her child $v_i$ to act in a similar
way, we say that the child node $v_i$ becomes ``activated" to the
message, otherwise ``inactivated". For each activated node $v_i$, we
assume all its activated parent nodes before $v_i$ influenced her,
and we define all these activated parent nodes in $P_i$ as activated
parent set $AP_i$ of node $v_i$.
\end{definition}

\begin{definition} [Information cascade]
Given network $G$ and a node being activated to a message $m$ at
$t_0$. The message cascades through the network with exposed child
nodes become activated at $t_1 > t_0$, thereby exposing their own
child nodes to the message, and so on. By repeating the process, an
information cascade is typically formed.
\end{definition}

\begin{definition} [Activation sequence]
Given network $G$ and an information cascade corresponding to a
message $m$, a set of nodes $\{ v_i \}$ capturing the order in which
the network nodes adopted the message $m$ is called the activation
sequence of the cascade.
\end{definition}

According to the above definitions, a information cascade can be
represented by an activation sequence as $ C_m = \{ v_1 ,v_2
,...,v_k  \}$. The time stamp that node $v_i$ gets activated to the
information is $t(v_i)$, $ t(v_i ) \le t(v_{i + 1} ) $. We denote
the time stamp sets corresponding to the activated parent set $AP_i$
and the activation sequence $ C_m$ as $APt_i$ and $ Ct_m $
respectively. Then the partially observed cascade before time $t$
can be denoted as $ C_m (t) = \{ v_i |t(v_i ) < t\}$, and the
cascade size can be defined as $|C_m (t) |$, where $|\cdot |$ is the
cardinality of a set. Moreover, we assume that the node which was
already activated to a message cannot be re-activated or inactivated
to the message.

\begin{definition} [History message set]
If node $v_i$ was activated to a message $m$, we say that message
$m$ is one of the history messages of node $v_i$. We explore history
message set $HM_i$ to represent all the history messages of node
$v_i$.
\end{definition}

\begin{definition} [Candidate message set]
If activated parent set $AP_i$ is not empty, node $v_i$ is exposed
to the messages published by the activated parent nodes in $AP_i$.
As $v_i$ is possible to be activated to the messages, we define the
messages as candidate message set $CM_i$ of $v_i$.
\end{definition}

\begin{problem} [Full-scale cascade dynamics prediction in
social networks] Given a network $G$, any time point $t$, partially
observed information cascade $C_m (t)$ before time $t$, the goal is
to propose a predictive framework based on $G$ and $C_m (t)$ such
that the framework can predict the activated node set $ ANSet_{t' }
$ and the cascade size $|ANSet_{t' }|$ at any later time $ t'
> t $ in the entire life of the cascade.
\end{problem}

\section{Full-Scale Cascade Dynamics Prediction Framework}\indent
In this section, we first introduce the framework to solve the
proposed full-scale cascade dynamics prediction problem in social
networks, and then explain the two main phases in the framework
respectively. Finally, we present FScaleCP properties and complexity
analysis.

\subsection{FScaleCP Framework}
To solve the challenges of full-scale cascade dynamics prediction,
we propose a framework, FScaleCP (shown in Figure 2), to first model
the local spreading behaviors, from which we then design a global
function for aggregating the local behavioral dynamics to
approximate the cascade dynamics in social network. Except for the
vivid way to cascades prediction, FScaleCP provides an intuitive and
comprehensive image to model the real-world network dynamics.

At the first phase, given network $G$ and history cascade set
$\{C_1, C_2, ..., C_n \}$, we propose a predictive model to estimate
the local spreading behaviors of inactivated nodes based on the
optimal feature space and model parameters. The motivation is that
when users spread messages, the exhibited behavioral patterns lead
to information redundancy that can be captured in terms of data
features. Following the machine learning research, we can learn an
estimation model by exploring a supervised learning framework that
utilizes the features.

At the second phase, based on the proposed spreading behavior
estimation model and partially observed cascade $C_m(t)$, we propose
an algorithm to predict the cascade conditions of message $m$ at
time $t'$. The idea is that the practical information cascading
process can be reproduced approximately by iterative propagation on
local network topology.

\begin{figure}[H] \small \centering
\includegraphics[width=8.5cm]{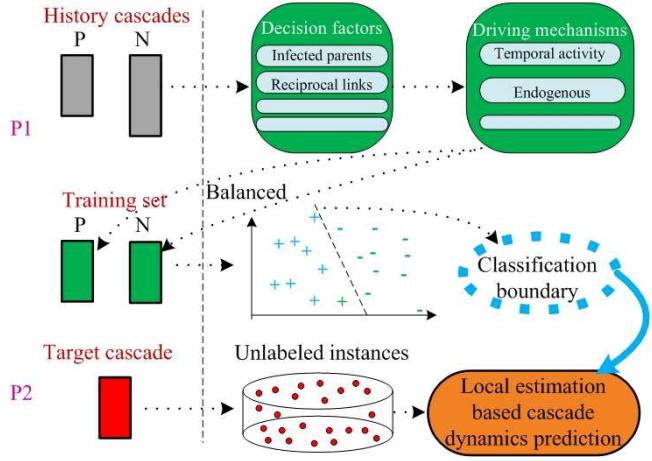}
\caption{Full-Scale Cascade Dynamics Prediction Framework
(FScaleCP).}\label{fig:2}
\end{figure}

\subsection{Local Behavioral Dynamics Estimation Model}

In this section, we assume a social media site's users have common
and consistent behavioral patterns, and they combine their personal
attributes and surrounding conditions to handle the candidate
messages. The information redundancies resulted by common and
consistent behavioral patterns can be utilized to identify their
behavior patterns in information diffusion process. Of course,
individuals can avoid such information redundancies by random their
response behaviors deliberately, such as selecting a completely
different responding behavior to similar messages and surrounding
conditions. Unfortunately, all of these requirements are contrary to
practical needs and human abilities. So this property can be
explored to help learn a spreading behavior estimation model.

\subsubsection{Local
Conditions Analysis and Driving Mechanism Definition}

In this study, we will analyze multiple driving mechanisms of
information diffusion to harness the redundant information, which
include (a) Content semantic driving mechanism (2) Temporal activity
driving mechanism (3) Network structural driving mechanism (4)
Endogenous driving mechanism.

\textbf{Content semantic driving mechanism.} Common sense dictates
that the candidate messages $CM_i$ and the history messages $HM_i$
all may influence users' responses in social networks. That is, the
content and the semanteme of the messages may drive users to spread
messages or not. In the candidate messages, keywords can help users
to identify core topics quickly, which can make messages avoid being
flooded by the continual stream of new messages. Moreover, more
longer the candidate messages, more information is contained in
them, and more possible to arouse users' interests to spread them.
In addition, users' history messages can hint for their personal
interest. More broad interest diversity of a user, more likely he
become interested in a random message and spread it. Finally, more
closely a candidate message's topic matches user's interest, more
possible the user spread it. Combining the above factors, we define
the content semantic mechanism $\mathcal{CSM}(t)$ at time $t$ as
follows:
\begin{equation}
{\mathcal{CSM}(t)} = f_{csm} (KeyW_m,ContLen_m,IntDiv_u,IntSim_m)
\end{equation}

\noindent where $ContLen_m$ is the length of a message, $KeyW_m$ is
a boolean value to denote the existence of keywords in message.
$IntDiv_u$ is the diversity of user interest computed by Shannon
entropy $IntDiv_u =  - \sum\nolimits_i {I_u (i)\log I_u (i)}$, where
$I_u (i)$ is the probability value of topic $i$ in user $u$'s
interest. $IntSim_m$ is the semantic similarity between user
interest and the candidate message computed by Jensen-Shannon
divergence $IntSim(I_u ,T_m ) = \frac{{D(I_u ||T_m ) + D(T_m ||I_u
)}} {2}$, where $ D(T_m ||I_u ) = \sum\nolimits_i {T_m (i)*In(T_m
(i))/I_u (i)} $ is the Kullback-Leibler divergence between
distribution $T_m$ and $I_u$, $T_m$ is the topic distribution of the
candidate message computed by LDA [27], $I_u$ is the average topic
distribution of all history messages of user $u$.

\textbf{Temporal activity driving mechanism.} The rationale for
taking a close look at this dimension is that messages are
time-sensitive, and their novelty and influence would decline with
the increase of time delay. Intuitively, for any appealing message,
most of users would likely to forward it as soon as possible to
improve their social influence. In other words, people may lose
interests to forward the messages that have been around for a long
time. Considering activation sequence $C_m$ and its time stamps
$Ct_m$, we define the temporal activity driving mechanism
$\mathcal{TAM}(t)$ at time $t$ as follows:
\begin{equation}
{\mathcal{TAM}(t)} = f_{tam} (AvgExpT_m,SurT_m,AvgFordD_m)
\end{equation}
\noindent where $AvgExpT_m$ is the average exposure time of
candidate message $m$ on $AP_i$, $SurT_m$ is the survival time
characterizing the novelty of message $m$, and $AvgFordD_m$ is the
average forward delay between every two successively activated nodes
for message $m$'s appealing degree measure.

\textbf{Surrounding conditions driving mechanism.} We notice that
the actions of friends would influence the user behaviors. According
to reality experience, close friends are always have much more
influence on users than ordinary friends. Moreover, the study in
[28] shows that repeated exposures have impact on information
spreading. Thus we study the influence of parents number,
relationship type and the ratio of activated parents on user
behaviors, and we define the surrounding conditions driving
mechanism $\mathcal{SCM}(t)$ as follows:

\begin{equation}
{\mathcal{SCM}(t)} = f_{scm} (SocialRe_u, SocialReR_u,
ActRecRel_u,ActRecRelR_u,RecRel_u,RecRelR_u)
\end{equation}

\noindent where $SocialRe_u$, $ActRecRel_u$ and $RecRelR_u$ are the
number of activated parents, reciprocal activated parents and
reciprocal parents respectively, $SocialReR_u$ is the ratio of
$SocialRe_u$  to the number of parents, $ActRecRelR_u$ is the ratio
of $ActRecRel_u$ to $SocialRe_u$, and $RecRelR_u$ is the ratio of
$RecRelR_u$ to the number of parents.

\textbf{Endogenous driving mechanism.} Recent research [25] suggests
the existing of limitation in human memory and cognition capability.
The results demonstrate that human's attention is limited. Thus, the
more number of the candidate messages, the less possibility of a
certain message being concerned. Moreover, users' attributes can be
viewed as the hints of user activeness. So we define the endogenous
driving mechanism $\mathcal{EM}(t)$ as follows:

\begin{equation}
{\mathcal{EM}(t)} = f_{em} (VerSta_u, ReMsg_u, ChlNode_u, AccCreT_u,
CPNodesR_u)
\end{equation}

\noindent where $VerSta_u$ is verification status of the account of
user $u$, $ReMsg_u$ is the number of related messages, $ChlNode_u$
is the number of child nodes, $AccCreT_u$ is the time when the
account being created, and $CPNodesR_u$ is the ratio of the the
number of child nodes to the number of parent nodes.

\subsubsection{Local Behavioral Dynamics Estimation}

Given network $G$ and current partially observed cascade $C_m(t)$,
we represent the Local Behavioral Dynamics Estimation Model as
$\mathcal{M}_{LBE}(G,C_m(t))$. Thus the model can be defined as a
function $f(\mathcal{CSM}(t))$, according to the content semantic
driving mechanism, $f(\mathcal{TAM}(t))$, according to the temporal
activity driving mechanism, and so on. We combine the four driving
mechanisms together to represent the final estimation model as
$\mathcal{M}_{LBE}(G,C_m(t)) = f(\mathcal{CSM}(t),
\mathcal{TAM}(t)), \mathcal{SCM}(t), \mathcal{EM}(t))$.

There are several ways to substantialize the function $f()$. In this
work, from a given network $G$ and a history cascade set $\{C_1,
C_2, ..., C_n \}$, each node may was activated to multiple messages.
We define each node related with different messages as multiple
different spreading instances. Thus we can get two disjoint sets of
instances: activated instances $P$ and inactivated instances $I$.
Based on the driving mechanisms $\{\mathcal{CSM}(t),
\mathcal{TAM}(t)), \mathcal{SCM}(t), \mathcal{EM}(t) \}$, we can get
a feature set ${\bf{x}} = [{\bf{x}}_{C}^T ,{\bf{x}}_{T}^T
,{\bf{x}}_{S}^T ,{\bf{x}}_{E}^T ]^T $ extracted from an instance.
Thus the activated instances $\mathcal{P}$ and inactivated instances
$\mathcal{I}$ can be represented by the feature sets as
$\mathcal{FP}$ and $\mathcal{FI}$. To differentiate the instances,
we define a new concept ``activation label", $y$, in this paper, to
show whether a node is activated or inactivated to a message. For a
given instance $n$, if $n$ is activated in the network, then $y(n) =
+1$, otherwise, $y(n) = -1$. As a result, we can have the
``activation labels" of instances in $\mathcal{FP}$ and
$\mathcal{FI}$ to be: $y(\mathcal{FP}) = +{\bf{1}}$ and
$y(\mathcal{FI}) = -{\bf{1}}$. By using $\mathcal{FP}$ and
$\mathcal{FI}$ as the positive and negative training sets, we can
build the Local Behavioral Dynamics Estimation Model
$\mathcal{M}_{LBE}(G,C_m(t))$ with a classification method, which
can be applied to predict whether an instance will be activated in
the network, i.e., the activation label of the instance. Let $n$ be
an instance to be predicted, by applying
$\mathcal{M}_{LBE}(G,C_m(t))$ to classify $n$, we can get the
activation probability of $n$ to be:

\begin{definition} [Activation Probability]
The probability that instance $n$'s activation states is predicted
to be active (i.e., $y(n) = +1$) is formally defined as the
activation probability of instance $n$: $p( y(n) = +1 | x(n))$,
where ${\bf{x(n)}} = [{\bf{x}}_{C}(n)^T ,{\bf{x}}_{T}(n)^T
,{\bf{x}}_{S}(n)^T ,{\bf{x}}_{E}(n)^T ]^T $.
\end{definition}

Due to the incomprehensibility of the training data, (i.e., linear
separable or not), in order to selecting a effective model, we
firstly compute the accuracies of multiple classifiers in candidate
model set $CM$, including supervised linear and nonlinear methods.
Then we acquire the linear separability of the training data by
comparing the accuracies. Moreover, as $\mathcal{M}_{LBE}(G,C_m(t))$
will be frequently conducted for cascade dynamics prediction, we
select a suitable model for local spreading behavior estimation
combining the model's complexity. We denote the process as $
ModelSelect(CM,\{\mathcal{FP}, \mathcal{FI}\}) $ in this paper.

Although many features are effective with regard to the spreading
behavior respectively (cf. Fig. 4), we find that the accuracy of the
proposed prediction method is not always increase with the number of
the available features (cf. Table 3). However, traditional methods
[1, 9, 10, 14, 26] simply consider all the available features as
indicator for information cascades prediction. To eliminate the
insignificant and redundancy features, we propose to conduct
Sequential Floating Backward Selection (SFBS) algorithm [28] among
the candidate features using the selected classifier as criterion
function to find the optimal feature space.

Understanding how the driving mechanisms influence the individual'
spreading behavior may potentially help us better model information
cascades. For this, the paper sums the weights of features belonging
to every driving mechanism to denote the indicative power of every
driving mechanism, which is denoted as $MecMeasure(Fw)$.

\renewcommand{\algorithmicrequire}{\textbf{Input:}}
\renewcommand{\algorithmicensure}{\textbf{Output:}}
\begin{algorithm}
\caption{Local Behavioral Dynamics Estimation Model Learning Method}               
\label{alg1}                         
\begin{algorithmic}[1]

\REQUIRE Network $G = (V,E)$, history cascades activation sequence set $\{C_1(t), C_2(t), ..., C_n(t) \}$.                  
\ENSURE Estimation model $\mathcal{M}_{LBE}(G,C_m(t))$, and driving mechanism measure vector $W$.    

\STATE  $\{ \mathcal{FP}, \mathcal{FI} \} \leftarrow
FeaturesSelection(G, \{C_1(t), C_2(t), ..., C_n(t) \})$ to get the
feature set of history cascade instances.

\STATE $M \leftarrow ModelSelect(CM,\{ \mathcal{FP}, \mathcal{FI}
\}) $ to select a suitable classifier for spreading behavior
estimation.

\STATE Set desired feature number: $i \leftarrow |\mathcal{FP}|/2$

\WHILE{$ i <= |\mathcal{FP}| $}

\STATE $Fr \leftarrow Cmp(Fr, SFBS(M, \{ \mathcal{FP}, \mathcal{FI}
\}, i) )$ to get the optimal feature space.

\STATE Update: $i \leftarrow i + 1$.

\ENDWHILE

\STATE $Fw \leftarrow M(Fr, \{ \mathcal{FP}, \mathcal{FI} \})$ to
get the feature weights.

\STATE Mechanism measure $W \leftarrow MecMeasure(Fw)$.

\STATE Return local spreading behavior estimation model
$\mathcal{M}_{LBE}(G,C_m(t))=\{M, Fr, Fw\}$, and mechanism measure
$W$.

\end{algorithmic}
\end{algorithm}

Based on the above analysis, The Pseudo code of Local Behavioral
Dynamics Estimation Model learning method is given in Algorithm 1.
Meanwhile, when applying the build model to predict instances in
independent instance set $\mathcal{IN}$ , the optimal labels
$\mathcal{Y}$, of $\mathcal{IN}$, should be those which maximize the
following activation probabilities:

\begin{equation}
\begin{split}
& {\mathop \mathcal{Y}\limits^ \wedge}_\mathcal{IN} = \arg \mathop {\max }\limits_{\mathcal{Y}_\mathcal{IN}} p(y(\mathcal{IN}) = \mathcal{Y}_\mathcal{IN}|G,C_m (t)) \\
& = \arg \mathop {\max }\limits_{\mathcal{Y}_\mathcal{IN}} p(y(\mathcal{IN}) = \mathcal{Y}_\mathcal{IN}|[{\bf{x}}_{C} (\mathcal{IN})^T ,{\bf{x}}_{T} (\mathcal{IN})^T ,{\bf{x}}_{S} (\mathcal{IN})^T ,{\bf{x}}_{E} (\mathcal{IN})^T ]^T ) \\
\end{split}
\end{equation}
\noindent where $y(\mathcal{IN}) = \mathcal{Y}_\mathcal{IN}$
represents that instances in $\mathcal{IN}$ have labels
$\mathcal{Y}_\mathcal{IN}$.

Given its importance in estimation model learning, we briefly
describe in more detail the SFBS algorithm. SFBS starts with the
whole feature subset and exclude features leading to the best
performance increase of the feature subset sequentially. Next, SFBS
includes one of the previously removed features if the resulting
subset would gain an increase in performance. This choice has been
made for the reasons that we assume most of the features are
effective and SFBS can investigate all possible feature
combinations. For clarity, following the original article [28], we
report the procedure of the SFBS in algorithm 2, which is the
expansion of step $\#5$ of Algorithm 1.

\renewcommand{\algorithmicrequire}{\textbf{Input:}}
\renewcommand{\algorithmicensure}{\textbf{Output:}}
\begin{algorithm}[h]
\caption{Sequential Floating Backward Selection (SFBS)}
\begin{algorithmic}[1]

\REQUIRE the set of all features, $Y = \{ y_1 ,y_2 ,...,y_d \}$, and the desired feature number.      
\ENSURE a subset of features, $X_k  = \{ x_j |j = 1,2,...,k;x_j  \in
Y\}$, where $0 \leqslant k \leqslant d$.   

\STATE Initialization: $X_k  = Y,k = d$

\STATE Step 1 (Exclusion):

\STATE $x^- = \text{ arg max } J(x_k - x), \text{ where } x \in X_k$

\IF{$J(x_k - x^-) > J(x_k):$}

\STATE $X_k-1 = X_k - x^-$

\STATE$k = k - 1$

\ENDIF

\STATE Go to Step 2

\STATE Step 2 (Conditional Inclusion):

\STATE $x^+ = \text{ arg max } J(x_k + x), \text{ where }  x \in Y -
X_k$

\STATE $X_k+1 = X_k + x^+$

\STATE $k = k + 1$

\STATE Go to Step 1

\STATE Termination: stop when k equals the number of desired
features.

\label{code:recentEnd}
\end{algorithmic}
\end{algorithm}

\subsection{Asynchronous Propagation based Cascades Dynamics
Prediction}

Propagation-based methods are fully distributed and localized. In
the propagation-based methods, each node can perform its operation
locally to achieve the global update over the whole network, without
global information or controller. Meanwhile, synchronous method
assumes all the nodes perform their local operations in a certain
order, while asynchronous method can relax the constraint by
allowing each node to perform its operation in any order as long as
each node is involved in the operations with nonzero probability.
Inspired by the work about label propagation algorithm [15], this
paper propose an asynchronous propagation based method (FScaleAPM)
to predict full-scale cascades dynamics, where the activation labels
correspond to community labels and the local spreading behavior
estimation correspond to the label selection mechanism. The idea
behind FScaleAPM is that full-scale cascade dynamics prediction can
be realized by asynchronous local activation label update using the
spreading behavior estimation model on local network topology. That
is, we quantify how a user is influenced by its parents and conduct
the process on the inactivated neighbors (i.e., susceptible nodes)
of activated nodes iteratively.

Framework FScaleCP proposed in this paper is a general cascade
dynamics prediction solution and can be applied to represent the
information cascade approximately. When it comes to a cascading
process $C_m(t)$, the optimal labels of susceptible nodes
$\mathcal{N}$ will be:

\begin{equation}
\begin{split}
& {\mathop \mathcal{Y}\limits^ \wedge}_\mathcal{N} = \arg \mathop {\max }\limits_{\mathcal{Y}_\mathcal{N}} p(y(\mathcal{N}) = \mathcal{Y}_\mathcal{N}|G,C_m (t)) \\
& = \arg \mathop {\max }\limits_{\mathcal{Y}_\mathcal{N}} p(y(\mathcal{N}) = \mathcal{Y}_\mathcal{N}|[{\bf{x}}_{C} (\mathcal{N})^T ,{\bf{x}}_{T} (\mathcal{N})^T ,{\bf{x}}_{S} (\mathcal{N})^T ,{\bf{x}}_{E} (\mathcal{N})^T ]^T ) \\
\end{split}
\end{equation}

If nodes are not related to each other, that is, any node's behavior
is not dependent on the behavior of any one of the others, we define
them as ``independent nodes". Otherwise, we define them as
``correlation nodes". To the ``independent nodes", we can estimate
multiple nodes' local behaviors simultaneously according to Eq. (5).
As the susceptible nodes $\mathcal{N}$ always includes ``correlation
nodes", the above target function is very complex to solve. In this
paper, to ``correlation nodes" $\{n_1, n_2, ..., n_k \}$, we propose
to obtain the labels of ``correlation nodes" by updating one node
and fix the others, alternatively with the following equation:
\begin{equation}
\begin{split}
\left\{ \begin{array}{l}
 ({\mathop y\limits^ \wedge}_{n_1})^\tau = \arg \mathop {\max }\limits_{y_{n_1}} p(  y(n_1) = y_{n_1} | G, C_m (t), ({\mathop y\limits^ \wedge}_{n_2})^{(\tau-1)}, ({\mathop y\limits^ \wedge}_{n_3})^{(\tau-1)}, ..., ({\mathop y\limits^ \wedge}_{n_k})^{(\tau-1)}  ) \\
 ({\mathop y\limits^ \wedge}_{n_2})^\tau = \arg \mathop {\max }\limits_{y_{n_2}} p(  y(n_2) = y_{n_2} | G, C_m (t), ({\mathop y\limits^ \wedge}_{n_1})^\tau, ({\mathop y\limits^ \wedge}_{n_3})^{(\tau-1)}, ..., , ({\mathop y\limits^ \wedge}_{n_k})^{(\tau-1)} ) \\
 \quad \quad \quad ...... \\
 ({\mathop y\limits^ \wedge}_{n_k})^\tau = \arg \mathop {\max }\limits_{y_{n_k}} p(  y(n_k) = y_{n_k} | G, C_m (t), ({\mathop y\limits^ \wedge}_{n_1})^\tau, ({\mathop y\limits^ \wedge}_{n_2})^\tau, ..., ({\mathop y\limits^ \wedge}_{n_{(k-1)}})^\tau ) \\
\end{array} \right.
\end{split}
\end{equation}

Based on the precondition that each node only has two states and
activated nodes cannot be re-activated or inactivated, this paper
selects node from ``correlation nodes" randomly using uniform
probability to approximate the alternative update. Thus the selected
node can be updated as one of the ``independent node". Figure 3
shows an example of the above method, in which FScaleADM would
estimate the activation labels of the nodes in the updated
independent node set \{0, 2, 3, 5, 8\} according to Eq. (5). Note
that, about ``correlation nodes", the unidirectional correlation can
be viewed as a simplification of the bidirectional correlation, such
as (7, 8) and (1, 2).

\begin{figure}[!h] \small \centering
\includegraphics[width=14.8cm]{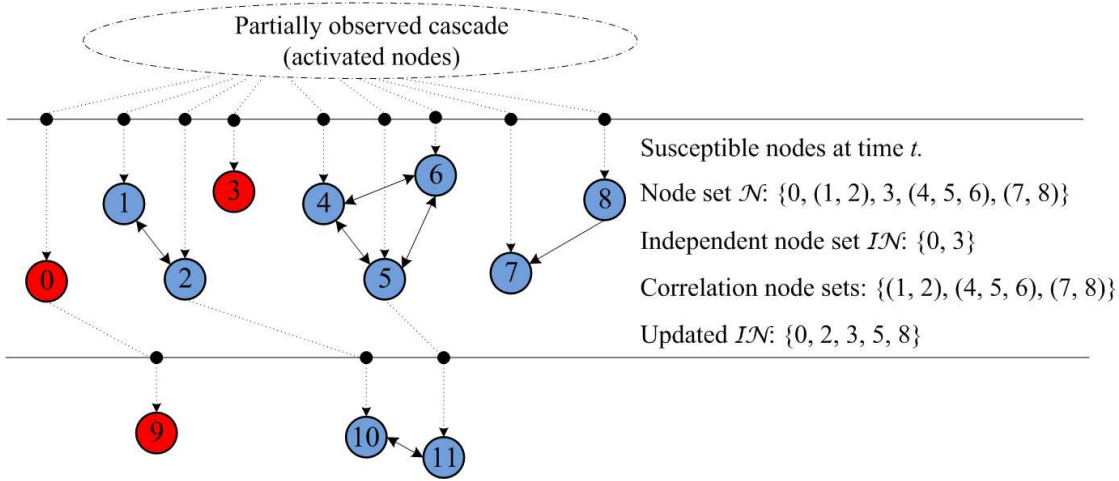}
\caption{The local topology example of information cascades for
FScaleADM. The topology contains two independent nodes and three
correlation node sets. The updated independent node set contains the
selected nodes from the correlation node sets after a random
selection. }\label{fig:2}
\end{figure}

In FScaleADM, the susceptible nodes $\mathcal{N}$ at time $t$ can be
viewed as a shell of the network. Ideally, the FScaleADM decomposes
a network into hierarchically ordered shells by recursively activate
the nodes with time stamps bigger than the ones of current shell. We
denote the time increment between shells as $\Delta T$ in this
paper. In reality, however, according to the discussion in the
previous paragraph, all the susceptible nodes $\mathcal{N}$ in each
shell are always cannot be updated and activated at one time but
rather be updated on multiple independent node sets step by step. We
define the time increment corresponding to the independent node sets
$\mathcal{IN}$ as $\Delta t$, $\Delta t =
|\mathcal{IN}|/|\mathcal{N}| \cdot \Delta T$. In this way, the
spreading behaviors of the nodes in network shell are estimated
asynchronously by asking all its neighbors' labels. The overview of
FScaleADM is presented in Algorithm 3.

\renewcommand{\algorithmicrequire}{\textbf{Input:}}
\renewcommand{\algorithmicensure}{\textbf{Output:}}
\begin{algorithm}
\caption{FScaleADP Method}               
\label{alg1}                         
\begin{algorithmic}[1]
\REQUIRE (a) social network with partially observed cascade, (b)
local spreading behavior estimation model $\mathcal{M}_{LBE}(G,C_m(t))=\{M, Fr, Fw\}$, (c) time increment $\Delta T$ and prediction time period $T$.        
\ENSURE Cascade condition after time period $T$.       

\STATE Get activation sequence $C_m$ and relevant time stamps $Ct_m$
from partially observed cascade

\STATE Find the newest time stamp $t_{new}$ from $Ct_m$ and set $t
\leftarrow t_{new}$.

\STATE Compute susceptible nodes $\mathcal{N}$ based on activation
sequence $C_m$ at current time.

\WHILE{$t <= t_{new} + T$}

\STATE ${\text{  }}$ (a) Get independent node set $\mathcal{IN}$
from susceptible nodes $\mathcal{N}$ with uniform probability..

\STATE ${\text{  }}$ (b) Conduct $\mathcal{M}_{LBE}(G,C_m(t))$ on
$\mathcal{IN}$ to update activation sequence $C_m$ and susceptible
nodes $\mathcal{N}$.

\STATE ${\text{  }}$ (c) Compute time increment: $\Delta t =
|\mathcal{IN}|/|\mathcal{N}| \cdot \Delta T$.

\STATE ${\text{  }}$ (d) Update system time: $t \leftarrow t +
\Delta t$.

\ENDWHILE

\STATE Return the predicted cascade condition at time $t_{new} + T$
based on activation sequence $C_m$.

\end{algorithmic}
\end{algorithm}

To illustrate FScaleADM Method, we will use a running example shown
in Figure 4 to provide a demonstration. Initially, at time $t$, the
susceptible node set $\mathcal{N}$ is \{0, (1, 2), 3, (4, 5, 6), (7,
8)\}, the independent node set $\mathcal{IN}$ is \{0, 3\} and
correlation node sets are \{(1, 2), (4, 5, 6), (7, 8)\}. FSclaeADM
selects nodes \{2, 5, 8\} from correlation node sets randomly and
add them into independent node set $\mathcal{N}$. Then it estimates
the spreading behaviors on the updated independent node set \{0, 2,
3, 5, 8\} with time increment ${5 \mathord{\left/
 {\vphantom {5 {9 \cdot }}} \right.
 \kern-\nulldelimiterspace} {9 \cdot }} \Delta T$. Then it activates the
nodes \{0, 2, 3, 5\} and updates the activated node set and
susceptible node set. At time $t + {5 \mathord{\left/
 {\vphantom {5 {9 \cdot }}} \right.
 \kern-\nulldelimiterspace} {9 \cdot }} \Delta T$, the
susceptible node set $\mathcal{N}$ is \{9, 1, (10, 11), (4, 6), (7,
8)\}, the independent node set $\mathcal{IN}$ is \{9, 1\} and the
updated independent node set $\mathcal{IN}$ is \{9, 1, 10, 4, 8\}.
Then FSclaeADM activates the nodes \{9, 1, 10, 8\} with time
increment ${5 \mathord{\left/
 {\vphantom {5 {8 \cdot }}} \right.
 \kern-\nulldelimiterspace} {8 \cdot }} \Delta T$. In the same way, FSclaeADM updates nodes' spreading behaviors iteratively in the later time.

\begin{figure}[!h] \small \centering
\includegraphics[width=15.5cm]{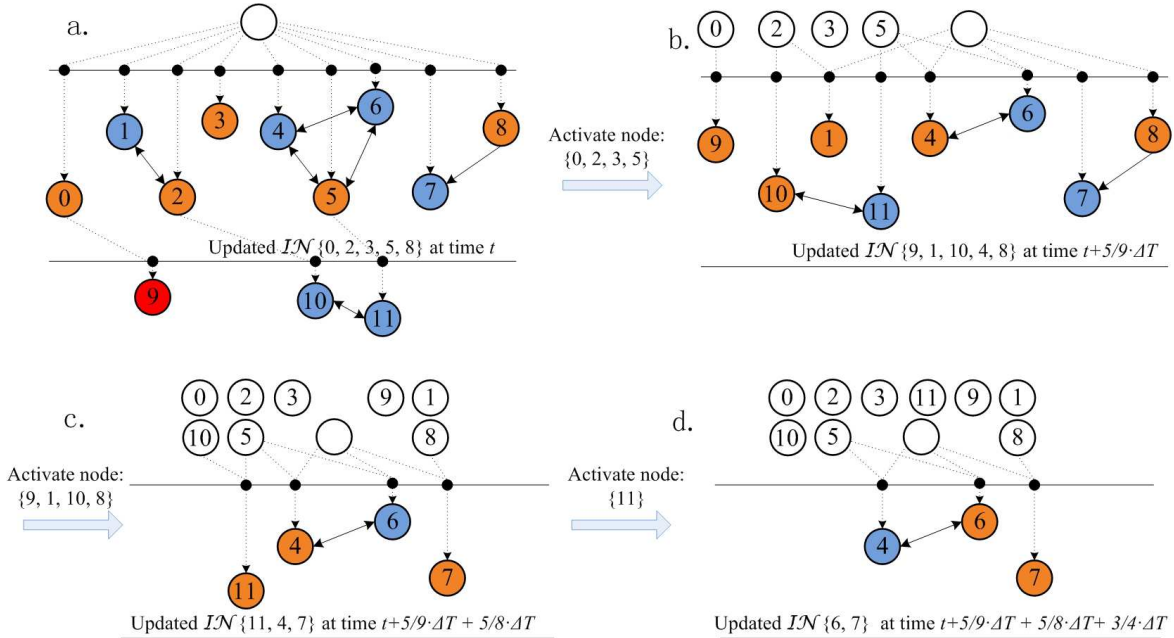}
\caption{Illustrate of FScaleADP method. To the cascade conditions
at different time, the nodes between two lines are the susceptible
nodes at current time, in which the ones with white color are the
activated nodes, the ones with orange color are the independent
nodes, and the ones with blue color are the correlation
nodes.}\label{fig:2}
\end{figure}

\subsection{FScaleCP Properties}
To prove the correctness and practicality of the FScaleCP framework,
we prove by induction its properties.

\begin{property} [Locality] Given a graph $G$, a cascade activation sequence
at time $t$  $C_m(t)$, then the information cascades prediction
$FScaleCP(G, C_m(t))$ equals to conduct asynchronous diffusion
process iteratively on the local topology of activated nodes, i.e.
$FScaleCP(G, C_m(t)) = FScaleADM(\mathcal{N}_t, C_m(t))$, where
$\mathcal{N}_t$ is the susceptible node set at time $t$.
\end{property}

As the messages propagate from activated nodes to inactivated nodes,
at every point of time, the asynchronous diffusion process will
merely be conducted on the topology region between the edge of the
activated community and the inactivated neighbors.

\begin{property} [Compositionality] Given a graph $G$, a cascade
activation sequence at time $t$  $C_t$ and the local topology of
activated nodes. Since local topology's bipartite graph structure,
there always has much independent nodes. To independent node set
$\mathcal{IN}$ in the local topology, consider any partition
$\mathcal{IN}_1$, $\mathcal{IN}_2$, $...$, $\mathcal{IN}_k$ of the
node set $\mathcal{IN}$, the asynchronous update process
$FScaleADP(\mathcal{IN}, C_t) = FScaleADP(\mathcal{IN}_1, C_t) \cup
FScaleADP(\mathcal{IN}_2, C_t) \cup ... \cup
FScaleADP(\mathcal{IN}_k, C_t)$.
\end{property}

This is a consequence of two facts: (1) Dividing the susceptible
nodes into multiple independent partitions with few interaction is
effortless, and (2) asynchronous diffusion process is conducted only
on the local network topology of the activated nodes.

\begin{property} [Convergence] Given a graph $G$, a cascade activation
sequence at time $t$ $C_t$,  then the cascading process would
converge if the local topology of activated users $G_t$ at multiple
consecutive time point are same, i.e. $G_{t_{i} } = G_{t_{i + 1} } =
G_{t_{i + 2} }$.
\end{property}
This is a consequence of the facts that each node has only two
states about a cascade and node state is irreversible. That is, the
cascading process is irreversible. So the convergence state must
exist even if $V(G_t) = \emptyset$.

Properties (1) and (2) have important computational repercussions.
As the size of cascades grows, storing complete cascade graph may
not be feasible. The locality property entails that FScaleCP can
adapt to any information cascade. Moreover, the locality property
and compositionality property entail that FScaleCP algorithm is
highly parallelizable, because it can estimate the node behavior
dynamics of different independent partitions simultaneously with
relatively small combination work.

\section{Experimental Results}\indent
In order to evaluate the performances and fully demonstrate the
advantages of the proposed framework, we conduct a series of
experiments on real-world data set, and the results of multiple
tasks are reported.

\subsection{Experimental Setup}\indent

\textbf{Data Preparation.} The social network we used in this study
was crawled from Sina Weibo, which, similar to Twitter, is the
largest microblogging network of china. The crawled way is
illustrated in [26]. The data set includes 1.7 million users and 4
billion following relationships between them. For each user, the
data set collects her 1000 most recent microblogs and all her
profiles. In addition, the data set has 300000 microblog diffusion
traces. We define the messages spreaded by user $v$ as positive
training instances and the activated patents' messages that are
never been spreaded by node $v$ as negative training instances. As
positive and negative instances are much unbalanced, we sample a
balanced training data with equal number of positive and negative
instances.

\textbf{Comparison Methods.} In order to show the efficiency of our
proposed cascade prediction method, we compare the prediction
results with the following baseline methods. First, we evaluate the
performance on local spreading behavior estimation and e use LRC-Q1,
LRC-Q2 as baselines. Then, since we are the first to investigate
full-scale cascade dynamics prediction problem based on local
behaviors estimation, no previous models can be adopted as direct
baselines. Here, we use LRC-Q1, LRC-Q2 as the local spreading
behavior estimation module of FScaleCP for full-scale cascade
dynamics prediction evaluation. In addition, we implement the
CG-CPred method as a baseline for final cascade size prediction
evaluation.

\emph{FScaleCP with different feature sets}.  Comparing the
performances of FScaleCP with different features sets will
demonstrate the efficiency of the proposed local spreading behavior
estimation model and the effectiveness of the optimal feature space.
Meanwhile, comparing the performances of FScaleCP with different
features sets can deepen the understanding about the effect of
different driving mechanisms.

\emph{LRC-Q1}. LRC-Q [26] is a typical implementation of individual
spreading behavior estimation. In LRC-Q, the prediction task depends
on a group of features from close friends in ego networks. LRC-Q
measures pairwise influence using the theory of random walk with
restart and structure influence by calculating the number of
circles. LRC-Q trains logistic regression classifier for spreading
behavior prediction. Let $S_v$ denote the collection of active
neighbors in $v$'s ego network $G_v^t$ and $p_{v_i }$ denote the
random walk probability from the active user $v_i$ to the given user
$v$. Then the pairwise influence in LRC-Q1 is the sum of the random
walk probabilities of all active neighbors, i.e.,
\begin{equation}
g(S_v ,G_v^t ) = \sum\limits_{v_i  \in S_v } {p_{v_i } }
\end{equation}
The structure influence in LRC-Q1 formulated by:
\begin{equation}
f(S_v ,G_v^t ) = e^{ - \mu |C(S_v )|}
\end{equation}
where $C(S_v )$ is the collection of circles formed by the active
neighbors and $\mu$ is a decay factor.

\emph{LRC-Q2}. Rather than using the definition of pairwise
influence and structure influence in Eq. (8) and (9). LRC-Q2
consider the influence of time and the number of active neighbors,
and it defines them as follows:
\begin{equation}
g(S_v ,G_v^t ) = \sum\limits_{v_i  \in S_v } {h_{v_i }p_{v_i } }
\end{equation}
\begin{equation}
f(S_v ,G_v^t ) = a\log (|S_v | + 1) + be^{ - \mu |C(S_v )|}
\end{equation}
where $h_{v_i }$ is the time difference and $a$ and $b$ are two
balance parameters.

\emph{Cascade graph based cascade prediction (CG-CPred)}. To
represent the methods utilizing the features generated from cascade
graphs in early stage, we design cascade prediction approach
CG-CPred capturing content semantic factors, temporal activity
factors and the structural features of the cascade graphs in Sina
Weibo scenario based on the method [14].

\textbf{Evaluation Metrics.} We perform 10-fold cross validation and
evaluate the performance of different approaches in term of
Precision (Prec.), Recall (Rec.), F1-measure (F1), and
Accuracy(Acc.).

In our evaluation, we set the time increment $\Delta T$ to be five
minutes and set the prediction period with different value according
to prediction task. A large time increment $\Delta T$ may lead to
network nodes miss the best spreading opportunity, and a small
$\Delta T$ may bring too much spreading behavior estimation
manipulations in the process of cascade dynamics prediction. So we
set the value of $\Delta T$ according to people's online behavior
habits. Observe that a large number of microblogs are popular for
only one day and a much large number of microblogs has never been
popular. To the popular microblogs, the contagion rate slows down
after two days when the microblogs have been published, and the
contagion duration of most of them are in the range of five days.

\subsection{Performance validation of FScaleCP}\indent

\subsubsection{Behavioral Dynamics Estimation and Driving Mechanism Measure}\indent

\textbf{Choice of Spreading Behavior Estimation Model.} In this
section, we consider the task of finding a suitable model for
spreading behavior estimation. Based on all available features, we
perform the classification task using a range of learning
techniques, including Logistic Regression, Naive Bayes, SVM,
Decision Tree and Random Forest. The results of them are summarized
in Table 1. By comparing the accuracies of the linear and nonlinear
methods, we can conclude that the training date is linear separable.
Moreover, Table 1 presents that Random Forest obtains the best
performance and Decision Tree gets the second-best performance in
the spreading behavior estimation task. However, random Forests is
more vulnerable to overfitting, hence, we select the Decision Tree
classifier as FScaleCP's spreading behavior estimation model. Except
for Naive Bayes, the similar results of the other methods show that
when sufficient information is available in features, the user
identification task becomes reasonably accurate and is not very
sensitive to the choice of learning algorithm.

\begin{table}[!h]\small \centering
\caption{The performances of all candidate classifiers in spreading
behavior estimation task.}
\includegraphics[width=6.0in]{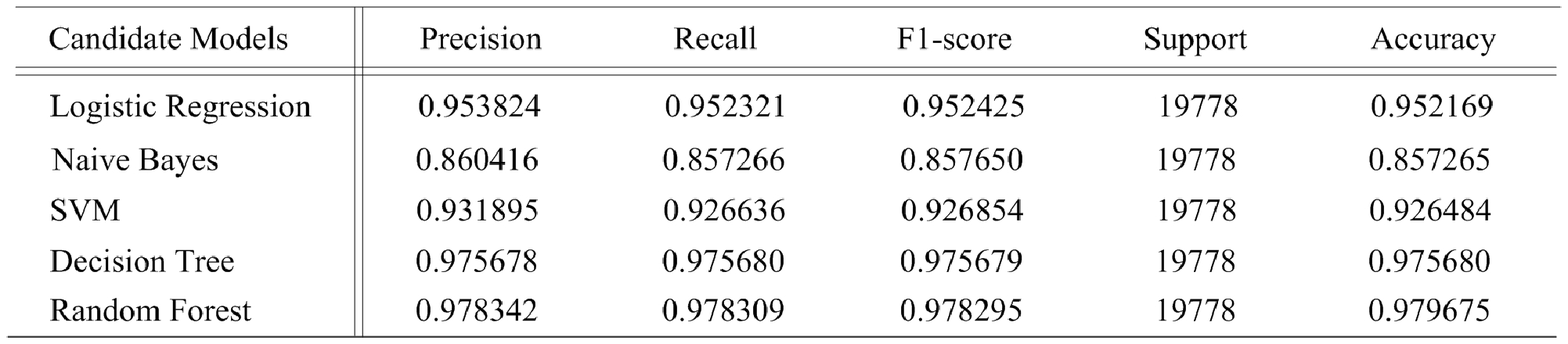}
\end{table}

\textbf{Choice of Feature Space and Driving Mechanism Measure.} To
show the correctness of the candidate features, we analyze the
effect of the features on the spreading behavior, and the results
are shown in Figure 5. In Figure 5 (a), we classify thousands of
messages according to their content length and count the number of
the messages being spreaded in the past, and we plot the spreading
probability of the messages under different content length. As the
content length increases, the spreading probability grows slowly. In
Figure 5 (b), we divide the data into four groups and find that the
messages including keywords get higher spreading probability in all
groups. Similarly, in Figure 5 (c), we can conclude that the
spreading probability increases along with the interest diversity
grows. Figure 5 (d) reports the average semantic similarity of
positive instances and negative instances in five group data. We can
find that positive instances have higher semantic similarity than
negative ones. In the same way, we report the spreading
probabilities under different average exposure time, average
forwarding delay, average survival time of messages, candidate
message number, social reinforcement and reciprocal active link
proportion in Figure (e), (f), (g), (h). We observe that spreading
probability is negatively correlated with average exposure time,
average forwarding delay, average survival time of messages and
candidate message number and positively correlated with social
reinforcement and reciprocal active link proportion. The above
results shows that the candidate features are effective in spreading
behavior estimation respectively.
\begin{figure}[!h] \small \centering
\includegraphics[width=5.5in]{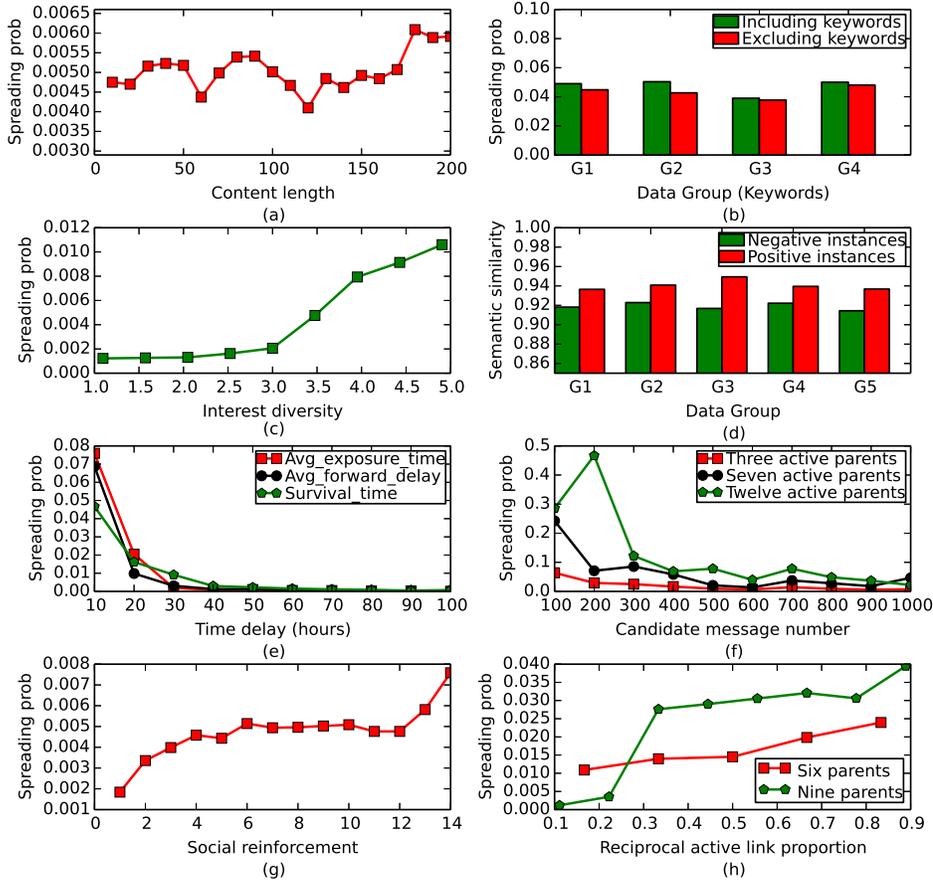}
\caption{Features correctness analysis. (a) Spreading probability
under different content length. (b) Influence of keywords on
spreading probability. (c) Spreading probability under different
interest diversity. (d) Semantic similarity of different instances.
(e) Spreading probability under different time delay. (f) Influence
of congnitive limitation on spreading probability. (g) Spreading
probability under different social reinforcement. (h) Spreading
probability under different reciprocal active link proportion.
}\label{fig:2}
\end{figure}

Although all the candidate features are impactful in the task of
spreading behavior estimation respectively, the predictive power of
some of them is negligible and there may have correlation between
them, which leads to unsatisfactory results and bad interpretability
of estimation model. In order to distinguish the key factors from
all the candidate features, we conduct the SFBS algorithm among them
using Decision Tree classifier as criterion function and identify
the optimal feature space from the 18-dimensional complete feature
space. The result is shown in the second column of Table 2 (a) in
descending order of weight.

\begin{table}[H]\small \centering
\caption{Feature evaluation and driving mechanism
measure.}\label{fig:side} \vspace{\baselineskip} \centering
\subtable[Considering survival time]{ \label{fig:side:a}
\includegraphics[width=3.0in]{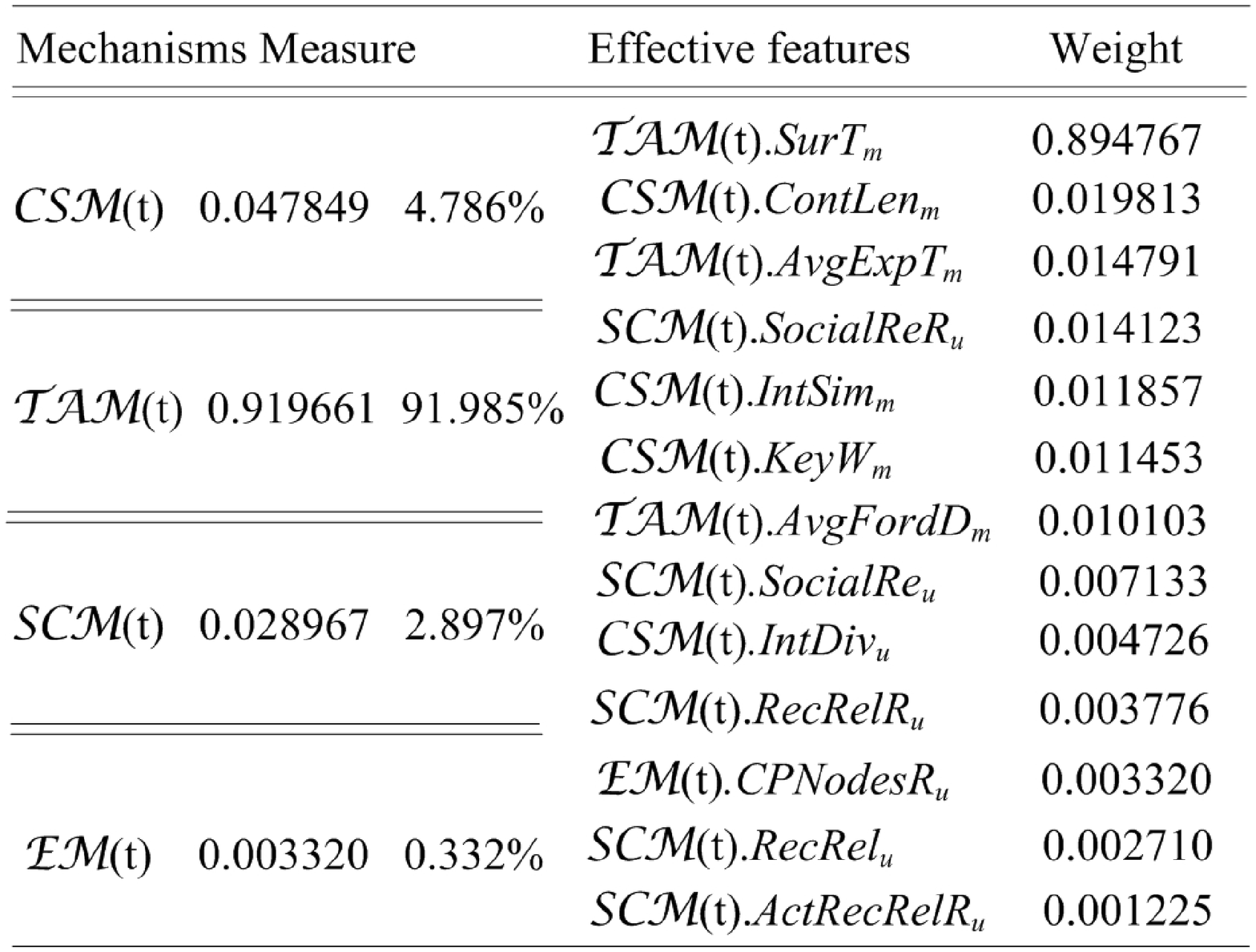}}
\hspace{8ex} \subfigure[Excluding survival time]{ \label{fig:side:b}
\includegraphics[width=3.0in]{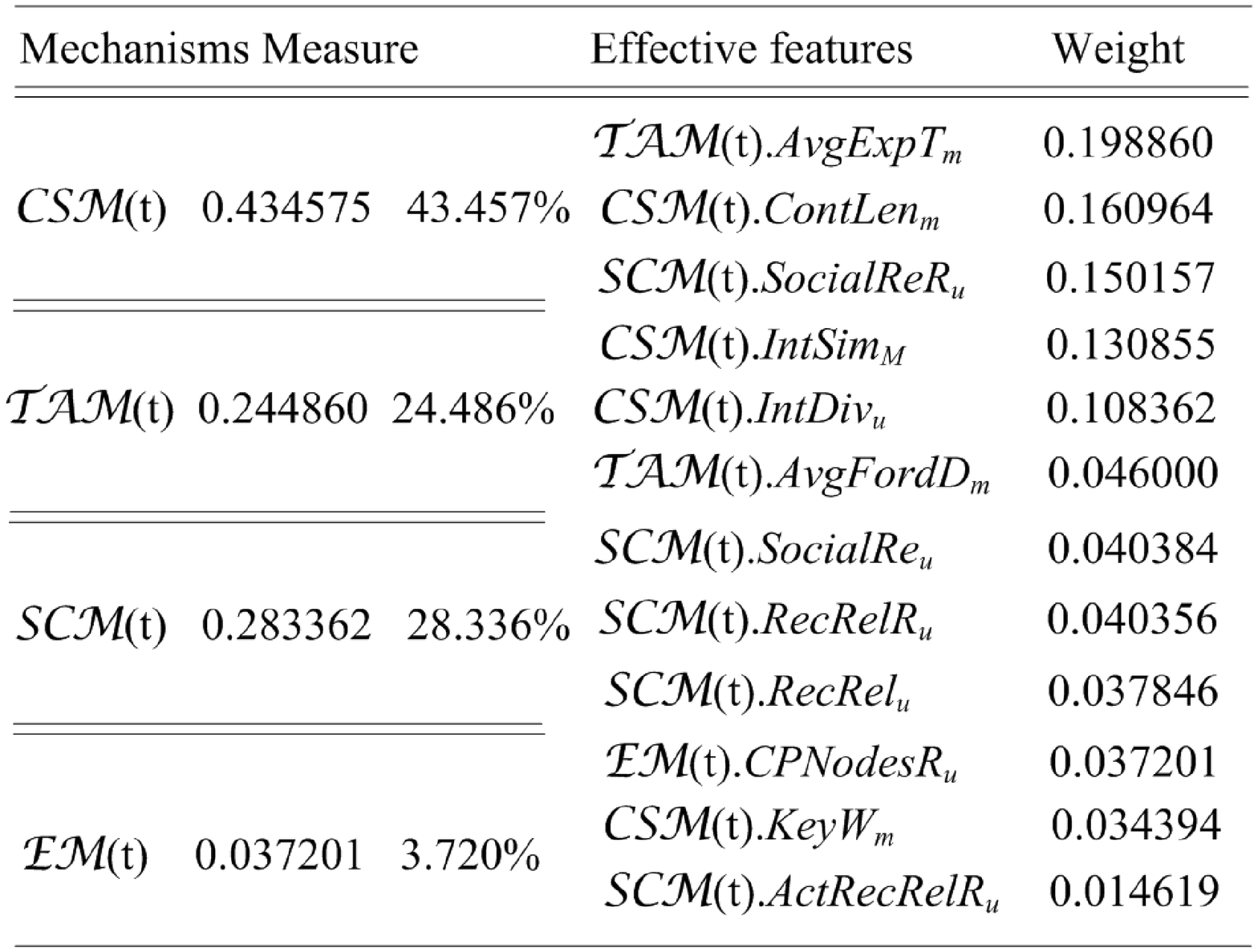}}
\end{table}

\begin{figure}[!h] \small \centering
\includegraphics[width=4.8in]{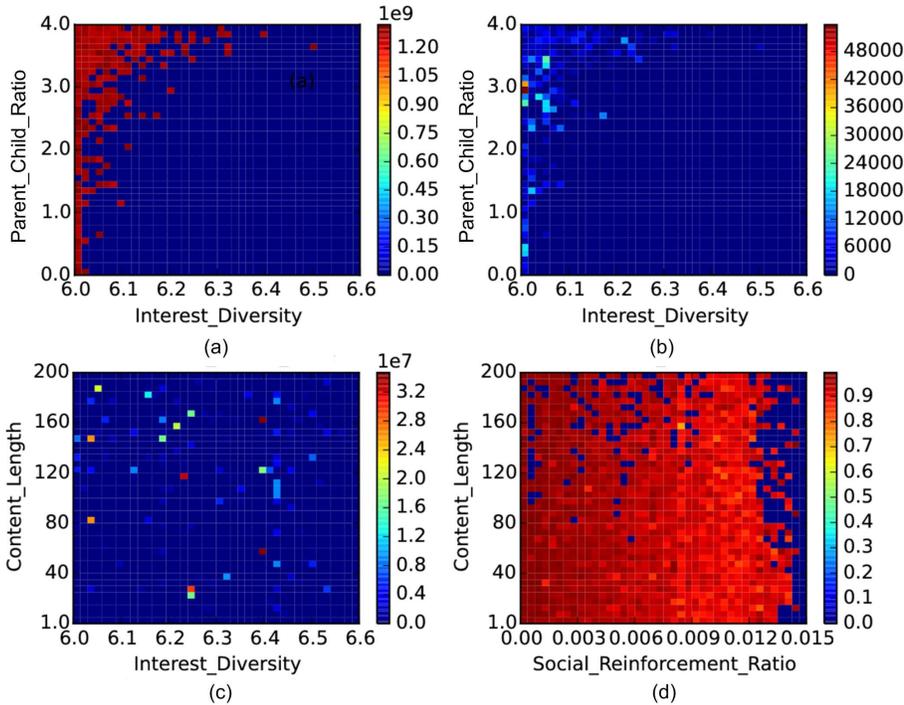}
\caption{Features correlation analysis. (a) Heatmap of account
creation time as a function of the ratio of child node number to
parent node number and interest diversity, (b) Heatmap of related
messages number as a function of the ratio of child node number to
parent node number and interest diversity, (c) Heatmap of average
exposure time as a function of content length and interest
diversity, (d) Heatmap of semantic similarity as a function of
content length and social reinforcement ratio.}\label{fig:2}
\end{figure}

According to Table 2 (a), the feature with the greatest weight is
survival time $SurT_m$ of messages, which show that the most
important factor for people in messages spreading is the novelty of
messages. The next most effective feature is content length
$ContLen_m$ of messages. Among the 13-dimensional optimal feature
space, 4 features are elements of content semantic driving mechanism
${\mathcal{CSM}(t)}$, 3 features are elements of temporal activity
driving mechanism ${\mathcal{TAM}(t)}$, 5 features are elements of
surrounding conditions driving mechanism ${\mathcal{SCM}(t)}$ and 1
feature is element of endogenous driving mechanism
$\mathcal{EM}(t)$. Based on the feature weights, the proportions of
the driving mechanisms are 4.786\%, 91.985\%, 2.897\% and 0.332\%
respectively. The result reveal a strong relationship between the
${\mathcal{TAM}(t)}$ and the spreading of messages. As the weight of
feature $SurT_m$ is much larger than the others, we analyze the
performance of the other features in detail to better understanding
the driving mechanisms of information cascades, and the result is
shown in Table 2 (b). We can find that the estimation accuracy
decreases from 0.977449 to 0.864353, and the most important feature
is the average exposure time $AvgExpT_m$ and the the most powerful
driving mechanism is content semantic driving mechanism
${\mathcal{CSM}(t)}$.

Moverover, we explain the necessity of feature selection from
correlation between features and spreading behavior estimation
accuracy. As shown in Figure 6 (a), (b), the account creation time
and related messages number have large values at the area of small
interest diversity and large ratio of child node number to parent
node number. Thus, the account creation time and related messages
number can be expressed by the two features. Meanwhile, Figure 6 (c)
show that average exposure time is not associated with content
length and interest diversity. Similarly, Figure 6 (d) show that
semantic similarity is not associated with content length and social
reinforcement ratio. Thus, the fact that average exposure time and
semantic similarity belong to the optimal feature space while
account creation time and related messages number do not can as a
exemplification to show that there have correlation between
candidate features and FScaleCP can identify independent and
effective features from them.

\textbf{Spreading Behavior Estimation.} Table 3 reports the
estimation results by different methods using different features.
From the table, we make the following observations. First, for
FScaleCP method, the optimal features lead to the best accuracy and
the complementary features generates the worst accuracy. The result
shows that FScaleCP is effective in feature selection. Second,
compared with LRC-Q1 and LRC-Q2, FScaleCP gets higher accuracy in
spreading behavior estimation. The result demonstrates that FScaleCP
has better performance than baseline methods. Third, compared with
FScaleCP using all candidate features, the better accuracy of
FScaleCP using optimal features shows that there may have conflict
between multiple effective features and the estimation accuracy is
not always increase along with the number of effective features.

\begin{table}[!h]\small \centering
\caption{Spreading behavior estimation result using different
features. OF: optimal features, AF: all candidate features, SF:
structural features, UA: user attributes, CF: complementary features
to OF.}
\includegraphics[width=6.0in]{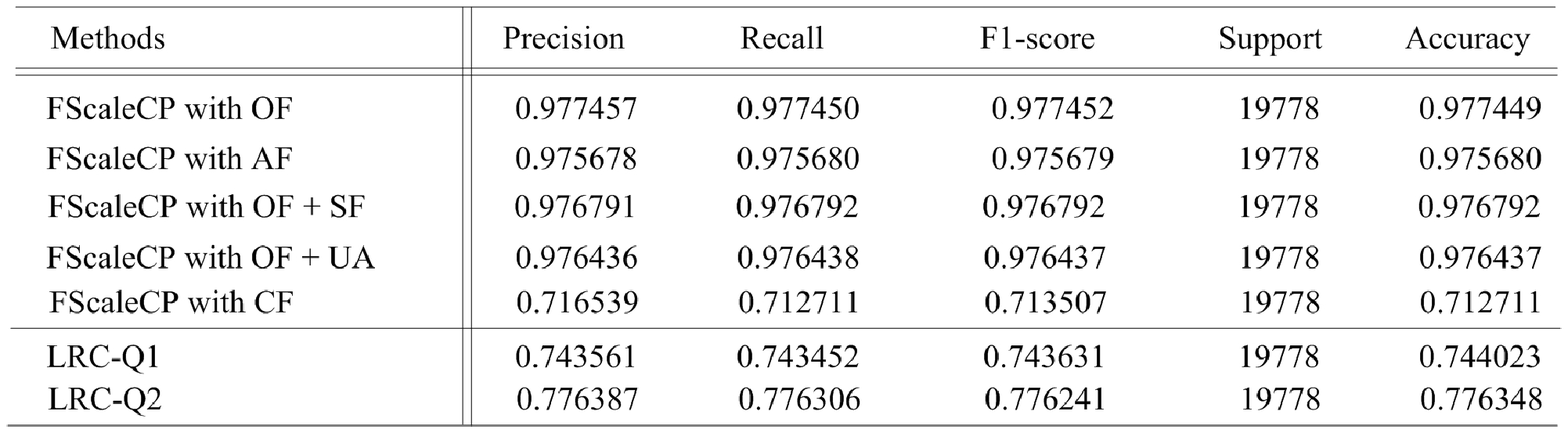}
\end{table}

\subsubsection{Full-Scale Cascade Dynamics Prediction}\indent

\begin{figure}[!h] \small \raggedleft
\includegraphics[width=7.0in]{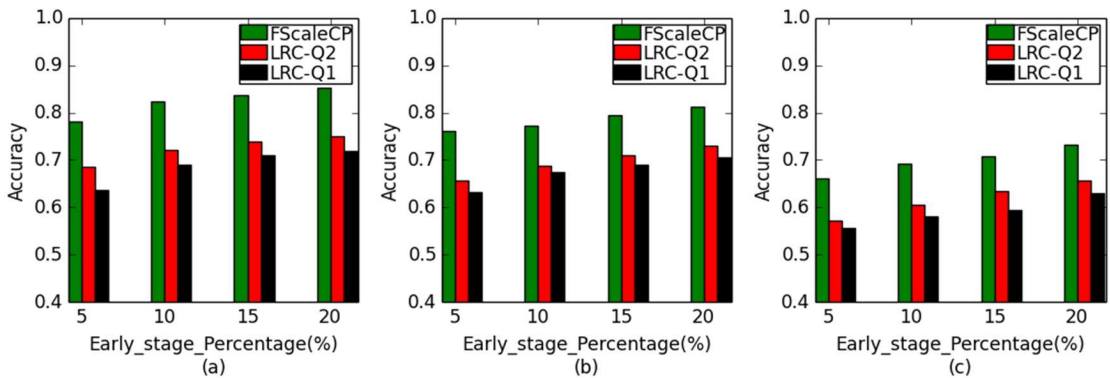}
\caption{Contagion states prediction on cascades with different
size. (a) Cascades with size at least 200. (b) Cascades with size at
least 400. (c) Cascades with size at least 600. }\label{fig:2}
\end{figure}

\textbf{Contagion States Prediction.} We apply FScaleCP on cascades
with different total size in five days and evaluation the
performance of FSCaleCP by the activation accuracy obtained by
averaging over 30 cascades in each group. The final results are
shown in Figure 7. It can be seen that the methods generate
different accuracy for varying sizes of believable activated nodes
(from 5 percent to 20 percent) and the proposed method FScaleCP
significantly outperforms other baselines in different sized
cascades. We can see that the accuracy value grows along with the
early stage percentage increases. Note that, the accuracy growth is
not because the performance improvement of local spreading behavior
estimation but because there are more activated nodes as believable
information sources in cascade prediction. Moreover, the accuracy
values decrease along with the size of cascades increases. The main
reason is that a larger information cascade always presages a larger
scope of information contagion and a larger number of hierarchical
shells of the contagion network. The inaccuracy of local behavior
estimations will be transmitted and amplified along with number of
hierarchical shells increases.


\textbf{Cascade Process Prediction.} Except for the prediction of
the final contagion state of network nodes, the prediction of
contagion state at different time point of cascade processes is an
important purpose of FScaleCP. We collect different sized cascades
that are popular for only 2 days and predict the temporal contagion
states with ten percent believable activated nodes at early stage.
At every time point, the accuracies are computed based on all the
contagion states of network nodes before current time. Then we
average the prediction accuracies for all cascades and show the
results in Figure 8. Here, we discover that FScaleCP always carries
out the best performances in cascade process prediction. Moreover,
the accuracies decrease with the prediction period increases. As
local behavior estimation based FScaleCP is not directly related to
prediction period, we argue that the declines in performance are
mainly derived from the expansion of cascade ranges.

\begin{figure}[!h] \small \raggedleft
\includegraphics[width=7.0in]{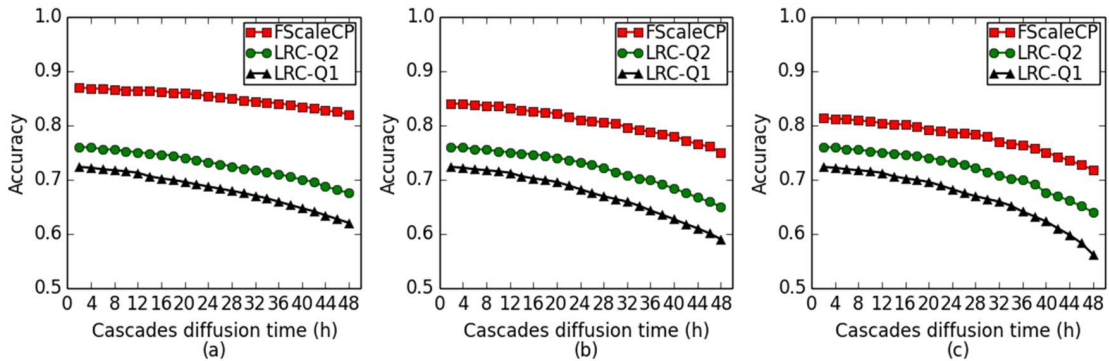}
\caption{Cascade process prediction (i.e. temporal contagion states
under different time point) on cascades with different size. (a)
Cascades with size at least 200. (b) Cascades with size at least
400. (c) Cascades with size at least 600. }\label{fig:2}
\end{figure}

\begin{figure}[!h] \small \raggedleft
\includegraphics[width=7.04in]{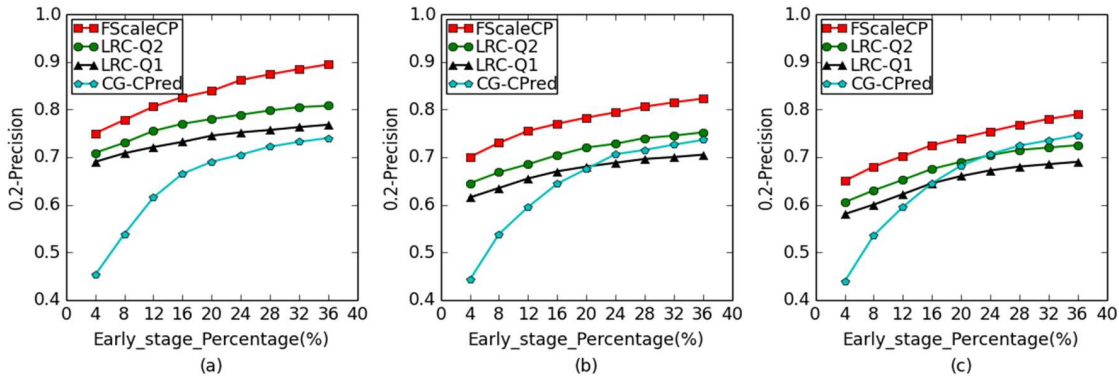}
\caption{Cascade size prediction on cascades with different size.
(a) Cascades with size at least 200. (b) Cascades with size at least
400. (c) Cascades with size at least 600. }\label{fig:2}
\end{figure}
\textbf{Cascade Size Prediction.} The final problem of this paper is
to predict the final size of cascades. For example, in the early
stage of a cascade, can we predict the final size of the cascade? We
evaluate the prediction performance with different percentage of
believable activated nodes in the cascades. In this case, we regard
the predicted value within the range of groundtruth twenty percent
as a correct prediction, and we use 0.2-Precision for result
evaluation. As shown in Figure 9, the FScaleCP method gets the best
performances on three groups of cascades in 0.2-Precision metric.
Moreover, a clear advantage of FScaleCP over traditional methods is
that FScaleCP is not dependent on the large percentage of early
stage, although it can improve FScaleCP's performance. We can see
that the precision of CG-CPred is unsatisfactory when only a small
percent of observed nodes available.

Compared with the traditional methods, the proposed method can
predict when and which nodes will be activated and is not dependent
on the percentage of observed nodes. However, we recognize that our
approach has prediction limitations in networks with large number of
hierarchical structure shells. To these networks, excellent
performance can be generated when multiple information sources are
scattered in networks.

\subsection{Complexity Analysis}

At every update period, the local spreading behavior estimation
model $\mathcal{M}_{LBE}(G,C_m(t))=\{M, Fr, Fw\}$ is conducted on
the nodes in independent node set $\mathcal{IN}$. The time
complexity is $O(|\mathcal{IN}|)$, where $|\mathcal{IN}|$ is the
size of the node set. To estimate the local spreading behaviors of
each node, the feature information of training instances is
required. Thus the time computation becomes $O(k \cdot
|\mathcal{IN}|)$, where $k$ is the number of features. During the
diffusion process in prediction period $T$, all independent node
sets need to be checked to estimate their nodes' spreading
behaviors, the approximate total check number is $T / \Delta t $,
where $\Delta t $ is the time increment corresponding to each
independent node set. Thus the time complexity is $O(k \cdot
|\mathcal{IN}| \cdot T / \Delta t)$. With the time increment $\Delta
T$ between susceptible node shells to express $\Delta t$, $\Delta t
= |\mathcal{IN}| / |\mathcal{N}| \cdot \Delta T$, the time
complexity can be rewritten as $O(k \cdot |\mathcal{IN}| \cdot T /
(|\mathcal{IN}| / |\mathcal{N}| \cdot \Delta T)) $ = $O(k \cdot
|\mathcal{N}| \cdot T / \Delta T) $. Note that, $|\mathcal{N}|$ is
not constant, and it varies along with the diffusion process.

\section{Discussion and Conclusion}\indent

Cascades prediction has always been an important research topic.
Though many related researches have been done, most of them focus
only on indicator features exploration and final size prediction,
and little has been down in problem of investigating cascade
dynamics comprehensively. In this paper, we have proposed a
local-first cascade dynamics prediction framework FScaleCP. The
proposed framework can predict cascade dynamics and is dependent of
the cascade graph in early stage. Moreover, FScaleCP is not only
predicting the final size of cascades, but also when and which nodes
will be activated. By the driving mechanisms measure, FScaleCP
identifies the most cardinal influencing factors and deepens the
understanding about cascade dynamics. The results would provide
basis for cascades control and theoretical modeling of information
cascades. Finally, experiments results show that the proposed
FScaleCP perform better than baseline methods.

Overall, FScaleCP is a practical yet general approach since it
mainly focuses on modeling the cascade dynamics. In this paper, we
just simply explore some basis influencing factors as features,
other various factors can be integrated into the framework
conveniently. Moveover, we assume that the network structure is
static in cascade process. However, the real-world network structure
varies along with the social interaction between uses. Therefore, it
is necessary to understand cascade dynamics using dynamic social
networks.

\section*{Acknowledgements}\indent
The work was supported partially by the National Natural Science
Foundation of China (Grant No. 61202255), University-Industry
Cooperation Projects of Guangdong Province (Grant No.
2012A090300001) and the Pre-research Project (Grant No.
51306050102). We thank Qicheng Zhang, Tao Zhou, JunMing Shao and
Duanbing Chen for their research and writing advices. We also thank
Yunpeng Xiao and Yuanping Zhang for careful reading of the
manuscript. The authors also wish to thank the anonymous reviewers
for their thorough review and highly appreciate their useful
comments and suggestions.

\section*{References}

\end{document}